\documentstyle[osa,manuscript,psfig]{revtex} 

\begin{document}
\titlepage

\title{Criticality in quark-gluon systems
far beyond thermal and chemical equilibrium} 

\author{Fu Jinghua$^1$, Meng Ta-chung$^{1,2}$, R. Rittel$^2$, 
and K. Tabelow$^2$ \protect\\
{\em $^1$Institute of Particle Physics, CCNU, 430079 Wuhan,
China}\\
{\em $^2$Institut f\"ur Theoretische Physik, FU-Berlin,
14195 Berlin, Germany}\protect\\
{\small e-mail: fujh@iopp.ccnu.edu.cn;
                meng@iopp.ccnu.edu.cn;
                meng@physik.fu-berlin.de;
                rittel@physik.fu-berlin.de;
                tabelow@physik.fu-berlin.de} }

\maketitle

\begin{abstract}
Experimental evidence and theoretical arguments for the existence of
self-organized criticality in systems of gluons and quarks are
presented. It is observed that the existing data for
high-transverse-momentum jet-production exhibit striking regularities;
and it is shown that, together with first-principle considerations,
such regularities can be used, not only to probe the possible
compositness of quarks, but also to obtain {\em direct evidence} for,
or against, the existence of critical temperature and/or critical
chemical potential in quark-gluon systems when hadrons are squeezed
together.
\end{abstract}


The question whether a ``new state of matter'', QGP (quark-gluon
plasma), can be created in laboratories has been raised soon after
physicists found that the fundamental building blocks of hadronic
matter are quarks and gluons, and that their (color) interactions can
be described by Quantumchromodynamics (QCD). Transitions between such
``new states'' and normal hadronic states are usually known 
\cite{CERN,jacob} as ``phase transitions'' in analogy to those
discussed in connection with critical phenomena in Equilibrium
Thermodynamics; and these transitions are expected to occur at a given
(critical) temperature and/or at a given (critical) chemical
potential. While a number of devoted high-energy nucleus-nucleus
collision experiments have been performed, no {\em direct signals} for
the existence of QGP has yet been found \cite{CERN}. 

With the help of high-energy proton-antiproton (Tevatron) and
heavy-ion (RHIC) colliders, it is now possible to measure
high-transverse-momentum ($p_T$), or -energy ($E_T$)-jets due to
``hard'' subhadronic scattering processes. It is anticipated
\cite{jacob,blazey} that such experiments would yield useful
information, not only on the structure of hadrons, but also on the
possible compositness of quarks \cite{blazey}. However, due to the
flexibility of the parton distribution functions (PDF's) it is
extremely difficult for experimentalists to draw definite conclusions
from their data \cite{blazey} --- independent of the accuracy of their
experiments.

In this paper: We report on {\em the observation of striking
regularities, in particular the observation of fingerprints of
self-organized criticality (SOC)}
\cite{BTWoriginal,BTWcontinue} in
hadron-induced and lepton-induced high-$E_T$ jet production
processes. We show that these observations can be used, not only in
probing the compositness of quarks, but also as {\em direct tests} for
the existence of thermal and/or chemical equilibrium, and thus as
direct tests for the existence of critical temperature and/or critical
chemical potential for quark-gluon systems 
\cite{CERN,jacob} when hadrons are squeezed together.
  
We recall \cite{letter}: The characteristic properties of the gluons
--- in particular the direct gluon-gluon coupling prescribed by the
QCD-Lagrangian, the confinement, and the non-conservation of
gluon-numbers, show that systems of interacting soft gluons should be
considered as open dynamical complex systems which are in general far
from thermal and/or chemical equilibrium. Taken together with the
observations \cite{BTWoriginal,BTWcontinue}
made by Bak, Tang and Wiesenfeld (BTW), this
strongly suggests the existence of self-organized criticality (SOC, i.e.
criticality without the need of fine tuning of any external parameter 
such as temperature) and thus the existence of BTW-avalanches in such
systems. 

We also recall: A small part of such BTW-avalanches manifest
themselves in form of {\em color-singlet} gluon clusters ($c_0^\star$)
which can be readily examined \cite{letter,longSOC}
experimentally in inelastic diffractive scattering processes. This is
because the interactions between the struck $c_0^\star$ and any other
{\em color-singlets} are of Van der Waal's type which are similar to
those between nucleons in nucleus (that is, they are much weaker than
color forces at distances of hadron-radius). In order to check the
existence and the properties of the $c_0^\star$'s, a systematic
data-analysis has been performed \cite{longSOC}, the result of
which shows that the distribution $D_S(S)$ of the size $S$, and the
distribution $D_T(T)$ of the lifetime $T$ of such $c_0^\star$'s {\em
indeed} exhibit power-law behaviors [namely $D_S(S)\propto S^{-\mu}$,
$D_T(T)\propto T^{-\nu}$, where $\mu$ and $\nu$ are positive real
constants].  Such characteristic features are known as {\em ``the
fingerprints of SOC''} \cite{BTWoriginal,BTWcontinue}.
In particular, taking into account, that in inelastic diffractive
scattering, the size $S$ of the struck $c_0^\star$ is proportional to
the directly measurable quantity $x_P$, which is the energy fraction
carried by ``the exchanged colorless object'' in such scattering
processes, it is found \cite{longSOC} from the existing data
\cite{HERAreviews} that $D_S(x_P)\propto
x_P^{-\mu}$, where $\mu =1.95\pm 0.12$, that is $\mu \approx 2$.

By considering inelastic diffractive scattering \cite{letter,longSOC},
we were able to check --- and {\em only} able to check (the existence
and the properties of) the {\em color-singlet} gluon-clusters.  Due to
the experimentally observed SU(3) color-symmetry, most of such
gluon-clusters are expected to carry color quantum numbers; that is,
they are color-multiplets (instead of singlets) which will hereafter
be denoted by $c^\star$'s.  But in accordance with the experimentally
confirmed characteristic features of the BTW theory, the existence of
SOC fingerprints in gluon systems should not depend on the dynamical
details of their interactions --- in particular the intrinsic quantum
numbers they exchange during the formation process.  This means in
particular, that the size ($S$) distribution $D_S(S)$ and the lifetime
($T$) distribution $D_T(T)$ of the $c^\star$'s are expected to have,
not only the same power-law behavior, but also the same power as those
of their color-singlet counterparts observed in diffractive scattering
processes \cite{letter,longSOC}!

Having in mind that quarks can be knocked-out of protons by
projectiles in deep-inelastic scattering processes \cite{HERAreviews}
where large momentum transfer between projectile and target takes
place, it is not difficult to imagine that colored gluon-clusters
($c^\star$'s) can also be ``knocked out'' of the mother proton ($p$)
by a projectile {\em provided that the corresponding transfer of
  momenta is large enough} (Here the knocked-out $c^\star$'s may or
may not have ``color lines'' connected to the remnant of the proton
--- depending on the final state of the struck $c^\star$ after it has
absorbed the object which delivers the momentum transfer).

In this paper, we show that there is indeed experimental evidence for
such knock-out processes. First, we point-out a striking regularity:
The differential cross section $d^2\sigma /dE_Td\eta$ (for $\eta
\approx 0$) data \cite{ppjet1a,ppjet1,ppjet1b,ppjet2} for inclusive
high-$E_T$-jet production processes $p+\bar{p}\rightarrow jet+X$
behaves like $E_T^{-\alpha}$, where $\alpha = 5,7$ and $9$ in the
lower, medium, and higher $E_T$-region, respectively \cite{DISremark},
as can be readily seen in the figure. Next we show that this behavior
is due to the probability distribution of the knocked-out colored
gluon clusters and the phase-space factors related to the processes in
which the large transverse momentum $p_T$ ($\approx E_T$) is
transferred.

We begin our discussion with the experimental observation
\cite{twojetdominance,jacob} that the inclusive high-$E_T$-jet
production process $p+\bar{p} \rightarrow jet+X$ is dominated by
two-jet-events.  Hence, it follows from the SOC-picture in which every
jet corresponds to a knocked-out colored gluon clusters $c^\star$,
that the invariant cross section $Ed^3\sigma /d p^3$ is expected to be
approximately proportional to the {\em square} of the probability
$D_S$ to find a $c^\star$.  The fact that the jet mass $m$ is
negligible small ($m^2 = E_T^2-p_T^2\approx 0$) implies that $E_T$ is
approximately the energy carried by the knocked-out $c^\star$.  Having
in mind that \cite{BTWoriginal,BTWcontinue,letter,longSOC} the size
$S$ of BTW-avalanches is directly proportional to the dissipative
energy it carries, we are immediately led to
\begin{equation}
\label{eq1}
D_S(E_T) \propto E_T^{-\mu}, \hspace{1cm} \mu \approx 2.
\end{equation}
Hence, we expect to see a factor $E_T^{-4}$ in the invariant cross section
$E d^3\sigma/dp^3$ which is related to the measured quantity shown
in the Figure as follows: $2\pi E_T^{-1} d^2\sigma/dE_T d\eta  
= E d^3\sigma/dp^3$.

We now show that the phase-space factors in $E d^3\sigma/dp^3$ which
corresponds to the different possibilities for a $c^\star$ to receive
a large transverse momentum $p_T$ ($\approx E_T$) are also powers in
$E_T$. In order to see this, we recall the well-known observation made
by Rutherford \cite{rutherford} in which the size and the location of
the effective scatterer in an atom is probed by examing
large-momentum-transfer scattering, and to recall a less-known
observation made by Williams \cite{williams} in which the following
has been pointed out: Ordinary space-time concepts are indeed very
useful for the semiclassical description of high-energy scattering
processes, provided that the de Broglie wave-length of the projectile
is short compared to the linear dimension of the scattering field,
{\em and} provided that the corresponding momentum-transfer which
determines the deflection is not smaller than the disturbance allowed
by the Uncertainty Principle. The above-mentioned phase-space factors
can be estimated by making use of the Uncertainty Principle and these
two observations \cite{rutherford,williams}.  To be more precise, when
we choose the $z$-axis as the collision axis and consider the
momentum-transfer in (or near) the $xy$-plane, the chance for two
objects (c.m. of $p$ and $\bar{p}$ and/or their constituents) moving
parallel to the $z$-axis to come so close to each other in the
$xy$-plane such that an amount of transverse momentum $p_T$ can be
transfered is approxiamtely proportional to the size of the
corresponding phase-space $\Delta x \Delta y \propto (\Delta p_x)^{-1}
(\Delta p_y)^{-1} \propto p_T^{-1} p_T^{-1} \propto E_T^{-2}$.  The
same procedure can be applied to the time-degree-of-freedom where we
obtain $\Delta t \propto (\Delta E)^{-1} \propto (\Delta E_T)^{-1}
\propto E_T^{-1}$, because not only a large momentum $p_T$ but also a
corresponding large energy ($E=E_T\approx p_T$) is transfered.

We recall \cite{HERAreviews,longSOC,letter} that, the proton $p$
(antiproton $\bar{p}$) is a spatially extended object with the
following preexisting constituents: (i) the valence quarks $q_v$
($\bar{q}_v$), which dominate the region $0.2 < x < 1$, where $x$
denotes the fraction of protons momentum they carry, (ii) the sea
quark-antiseaquark pairs $q_s \bar{q}_s$, which dominate the $10^{-2}
< x < 0.2$ region, and (iii) the gluon-clusters $c^\star$'s, which
dominate the $x < 10^{-2}$ region.  Hence, energy- and
momentum-conservation, and the above-mentioned properties of these
constituents dictate that we should see three different kinds of
processes which dominate three distinct kinematical regions
characterized by the values of $E_T$:

({\bf a}) In the low-$E_T$-region the dominating
contribution comes from One-Step (OS) processes
in which a $c^\star$ of size $S \propto E_T$ in $p$ collides with a
$c^\star$ of size $S \propto E_T$ in $\bar{p}$, where a transverse
momentum of the order $p_T$ is exchanged in the $c^\star$-$c^\star$
scattering.

({\bf b}) In the medium-$E_T$-region the dominating contribution comes
from Two-Step (TS) processes in which at least one of the quarks or
antiquarks from the sea is involved in the first step. We call
hereafter such processes TSS-processes. [Because of the obvious
symmetry in $p$ and $\bar{p}$, we discuss here only $c^\star$ in
$\bar{p}$ and $q_s$ (or $\bar{q}_s$) in $p$, but not the other
possible combination which also contributes.]  TSS-processes can take
place in two different ways (cf. Table): ({\bf b1}) A $c^\star$ of
size $S \propto E_T$ in $\bar{p}$ receives a transverse momentum
transfer of order $p_T$ from a $q_s$ of $p$ in the first step and
manifests itself as a high-$E_T$-jet.  In the second step the final
state of the scattered $q_s$ which is in general space-like,
encounters one of the $c^\star$'s in its surroundings transfering its
transverse momentum of order $p_T$, and thus give birth to the second
(approximately zero-mass) high-$E_T$-jet.  ({\bf b2}) A $q_s$ in $p$
meets a $q_s$ in $\bar{p}$, exchange a transverse momentum of order
$p_T$, and each of the scattered objects encounters a corresponding
$c^\star$ and thus give birth to two (approximately zero-mass)
high-$E_T$-jets.

({\bf c}) In the high-$E_T$-region the dominating contribution comes
from Two-Step (TS) processes in which at least one of the available
valence quarks or antiquarks are involved in the first step. We call
hereafter such processes TSV processes, which can also take place in
two different ways: ({\bf c1}) A $c^\star$ of size $S\propto E_T$ in
$\bar{p}$ receives a transverse momentum-transfer of the order $p_T$
from a $q_v$ in $p$ in the first step, and thus manifest itself as a
high-$E_T$-jet. In the second step, the scattered final state $q_v$
(which is in general space-like) encounters a $c^\star$ of size $S
\propto E_T$ transfering its transverse momentum of order $p_T$ and
become the second high-$E_T$-jet.  ({\bf c2}) A $q_v$ in $p$ meets a
$\bar{q}_v$ (or $q_s$ or $\bar{q}_s$) in $\bar{p}$, exchange $p_T$
($\approx E_T$) in the first step, and each of the scattered objects
encounters a suitable $c^\star$ in its surrounding and thus give birth
to two high-$E_T$ jets.

Note that
in the quantitative estimation of
the phase-space-factors the following empirical facts 
\cite{HERAreviews,longSOC,letter} as well as their logical
consequences have to be taken into account.

First, the $q_s$'s, the $\bar{q}_s$'s, and the gluons organized in
form of $c^\star$'s can be found everywhere inside the $p / \bar{p}$.
Hence the chance for a $c^\star$ (or $q_s$ or $\bar{q}_s$) of $p$ to
meet a $c^\star$ (or $q_s$ or $\bar{q}_s$) of $\bar{p}$ in a cell of
the size $\Delta x \Delta y$ should be the same as that for the c.m.
of $p$ and that of $\bar{p}$ to be in a phase-space-cell of the same
size. The size of this two-dimensional phase space is $\Delta x \Delta
y \propto p_T^{-2}\propto E_T^{-2}$.

Second, Two-Step processes (both TSS and TSV) can take place only when
there are suitable $c^\star$'s in the surroundings immediately after
the first step. The probability for having sufficient $c^\star$'s
around, whenever and wherever two constituents meet during the
$\bar{p}p$-collision is guaranteed when the c.m. of $p$ meets that of
$\bar{p}$. Hence, we can use the latter condition to perform the
estimation.  This means, the phase-space factor in the time-dimension
for TSS or TSV to take place is $(\Delta t)^2 \propto E_T^{-2}$.

Third, since there are only three $q_v$'s in $p$, and they can be
anywhere inside $p$, the chance for a $q_v$ to take part in the first
step of a TS process (i.e. for TSV) to occur is expected to be smaller
than that in TSS processes. Here, we need to find for example the
probability for a $q_v$ in $p$ to meet either any $q_s$ or any
$\bar{q}_s$ or one of the three $\bar{q}_v$'s in $\bar{p}$. By using
the same method, we readily see that it implies that an extra
phase-space-factor $\Delta x \Delta y \propto E_T^{-2}$ should be
taken into account.

The result of this estimation, summarized in the Table, explicitly
shows that the phase-space-factors for such high-$E_T$-jet production
processes can always be written as $E_T^{-\varphi}$ where $\varphi=2$,
$4$ and $6$ for OS and TSS and TSV processes respectively.
Furthermore, since the phase-space-factors and the probability for
obtaining $c^\star$'s of a given size are independent of one another,
we are immediately led to the conclusion
\begin{equation}
\label{eq2}
\frac{d^2\sigma}{dE_Td\eta} \propto E_T^{-\alpha},
\end{equation}
where $\alpha = \varphi -1 +2 \mu$, which is $5, 7,$ and $9$,
respectively (for $\mu \approx 2$).

Comparison with the obtained result given in Eq.(\ref{eq2}) and the
existing data \cite{ppjet1a,ppjet1,ppjet1b,ppjet2} is shown in the
Figure.

In conclusion, the results presented in Eqs.(\ref{eq1}) and
(\ref{eq2}) and in the Figure show that high-energy transverse-energy
($E_T$) jet-production experiments can be used to 
find the solutions of the following
longstanding puzzels: 

(I) The observation \cite{BTWoriginal,BTWcontinue} that self-organized
criticality (SOC) proposed by Bak, Tang and Wiesenfeld (BTW) can be
found in complex systems as diverse as earthquakes, sandpiles,
biological evolution and even stock markets, have triggered the
question: ``How general and/or how basic is SOC?''; in particular,
``Is this kind of behavior observed in open complex systems so
universal and so basic that SOC should exist also {\em at the
  fundamental level of matter} --- in the world of quarks and gluons
where the interactions are color forces prescribed by QCD?''  While
experimental indications for the existence of color-singlet
gluon-clusters in form of BTW-avalanches have already been found
\cite{letter,longSOC}, the existence of their colored counterparts ---
without which {\em the universality} of SOC at the fundamental level
of matter and the independence of dynamical details which is
intimately related to {\em the robustness} of its occurrence could not
be demonstrated --- remains to be shown. With the results summarized
in Eqs. (\ref{eq1}) and (\ref{eq2}), and in the Figure, we can now
answer the above-mentioned questions in the {\em affirmative}.

(II) Large-momentum-transfer scattering has been very helpful in
studying the structure of atoms \cite{rutherford} and that of nucleons
\cite{HERAreviews}; this idea of Rutherford \cite{rutherford} has also
been applied to probe the {\em possible compositness of quarks}
\cite{blazey}.  In fact, the existing data
\cite{ppjet1a,ppjet1,ppjet1b,ppjet2} for high-$E_T$-jet-production
have been compared \cite{blazey,ppjet1b,ppjet2} with the results
obtained from pQCD and the existing PDF's \cite{jacob,blazey}.  Due to
the flexibility of the PDF's, however, the question ``Is there {\em
  experimental indication for compositness} because the data exhibit a
behavior which differs from the predictions made on the basis of the
present theoretical understanding?'' could not be answered in a unique
way.  In this paper, we report on the observation of a striking
regularity, and we show that the differential cross section $d^2\sigma
/dE_T d\eta $ for $p+\bar{p} \rightarrow jet + X$ near $\eta =0$ is
expected to exhibit power-law behavior in $E_T$ in three distinct
$E_T$-regions with three different exponents, namely: $-5$, $-7$, and
$-9$. While the dynamical origin of such power-law behavior is SOC,
the powers ($\varphi$'s) in the phase-space-factors can be directly
derived from first principles. That is to say, comparison between the
existing data \cite{ppjet1a,ppjet1,ppjet1b,ppjet2} and the results
obtained from an analysis of the space-time properties of the
colliding objects have led us to the following conclusion: The
existing $p+\bar{p} \rightarrow jet + X$ data
\cite{ppjet1a,ppjet1,ppjet1b,ppjet2} {\em do not exhibit any
  irregularity} which could be considered as an indication for the
compositness of quarks.

(III) In the search for ``new state of matter'', concepts and methods
of conventional Equilibrium Thermodynamics play 
an extremely important role. In fact, it has always been {\em
taken for granted} that there is thermal and chemical equilibrium in
the relevant systems, and thus concepts such as temperature and
chemical potential are suitable quantities for the description of the
expected ``new state of matter QGP'' \cite{CERN,jacob}.
It seems that questions such as the following have never been
asked: What {\em can} we do, or {\em should} we do, if systems of
quarks and gluons are {\em far beyond} thermal and/or chemical
equilibrium?  Is there {\em direct} experimental evidence {\em for} or
{\em against} the existence of thermal and/or chemical equilibrium in
squeezed nucleon systems?''  Based on the results presented in this
paper, and the well-known fact (See e.g.
\cite{BTWoriginal,BTWcontinue,letter,longSOC}) 
that SOC can only exist in complex systems {\em far beyond}
equilibrium, we think it would be very useful to measure the
$E_T$-distributions of ($\eta \approx 0$) jet in high-energy
hadron-nucleus and nucleus-nucleus collisions, and {\em check whether
SOC-fingerprints also exist when more than two hadrons are squeezed
together}.

We thank F. Golcher, D. H. E. Gross, and W. Zhu for discussions and J.
Terron for correspondence.


\begin{figure*}
\rule{0cm}{0cm}\hfill\psfig{figure=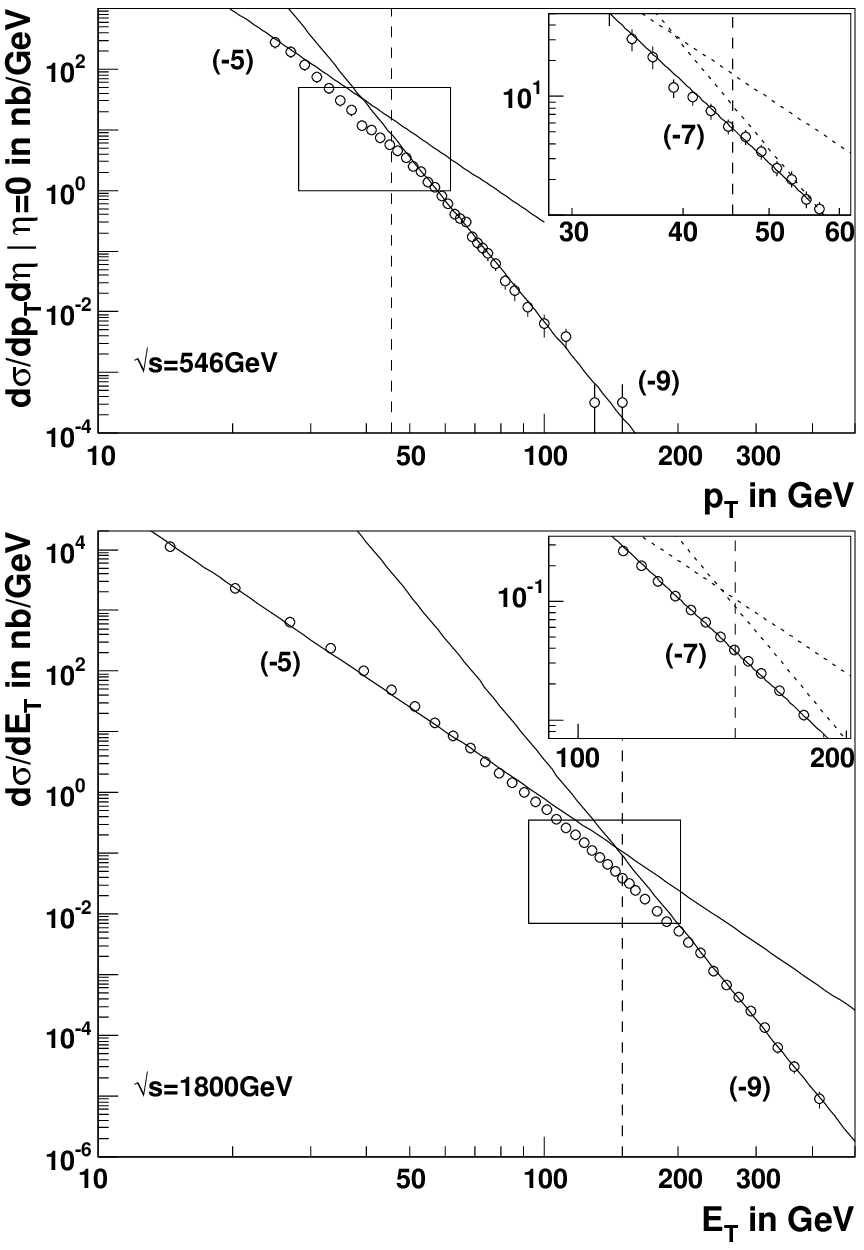}\hfill\rule{0cm}{0cm}
\caption{\label{fig1}
  Evidence for the validity of Eq.(\ref{eq2}) for
  $\bar{p}+p\rightarrow jet+X$ at $\eta\approx 0$ and different c.m.s
  energies $\sqrt{s}$.  Comparisons have been made with data in
  Refs.[\ref{ppjet1a}-\ref{ppjet2}]. For the sake of clearness, at
  $546$ GeV, only data in Ref.[\ref{ppjet1}] are shown in this figure.
  Those in Refs.[\ref{ppjet1a},\ref{ppjet1b}] and those at $630$ GeV
  [\ref{ppjet1}] exhibit the same regularities. The numbers shown in
  the brackets stand for the slopes of the straight lines in the
  log-log plots, that is, the value of $-\alpha$ in Eq.(\ref{eq2}).
  The vertical dashed line indicates $E_T = \sqrt{s}/12$.  We note:
  The empirical fact that there are three $q_v$'s in $p$, and about
  $50\%$ of the total c.m.s.  energy $\sqrt{s}/2$ of $p$ ($\bar{p}$)
  is carried by the gluons, suggest that TSV processes should begin to
  dominate for $E_T > \sqrt{s}/12$.}
\end{figure*}

\begin{table}
\rule{0cm}{0cm}\hfill\psfig{figure=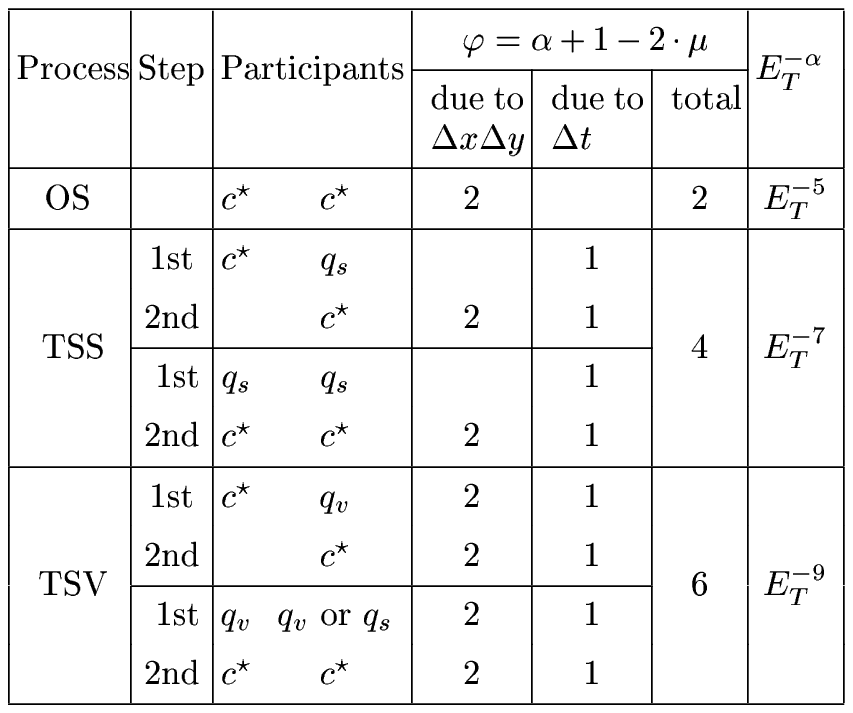}\hfill\rule{0cm}{0cm}

\vspace*{0.5cm}

\caption{\label{tab1} Estimation of phase-space-factors for different
processes by making use of the Uncertainty Principle and the
observations made in [\ref{rutherford},\ref{williams}]. See text for more
details. } 
\end{table}

\end{document}